\documentclass[conference]{IEEEtran}
\usepackage[latin9]{inputenc}
\usepackage{array}
\usepackage{verbatim}
\usepackage{rotating}
\usepackage{url}
\usepackage{multirow}
\usepackage{amsmath}
\usepackage{amssymb}
\usepackage{graphicx}
\usepackage{esint}
\usepackage[export]{adjustbox}

\makeatletter

\DeclareTextSymbolDefault{\textquotedbl}{T1}

\IEEEoverridecommandlockouts
\usepackage{cite}
\usepackage{amsfonts}
\usepackage{algorithmic}
\usepackage{textcomp}
\usepackage{xcolor}
\def\BibTeX{{\rm B\kern-.05em{\sc i\kern-.025em b}\kern-.08em
    T\kern-.1667em\lower.7ex\hbox{E}\kern-.125emX}}

\makeatother

\begin{document}
\title{Full-wave EM simulation analysis of human body blockage by dense 2D antenna arrays\thanks{Funded by the European Union. Views and
opinions expressed are however those of the author(s) only and do
not necessarily reflect those of the European Union or European Innovation
Council and SMEs Executive Agency (EISMEA). Neither the European Union
nor the granting authority can be held responsible for them. Grant
Agreement No: 101099491. Project HOLDEN.}}
\author{\IEEEauthorblockN{Federica Fieramosca\textit{$^{1}$}, Vittorio Rampa\textit{$^{2}$},
Michele D'Amico\textit{$^{1}$, }Stefano Savazzi\textit{$^{2}$}} \IEEEauthorblockA{\textit{$^{1}$}\textit{\emph{ }}DEIB, Politecnico di Milano, Piazza Leonardo da Vinci 32, I-20133, Milano, Italy\textit{}\\
\textit{$^{2}$}\textit{\emph{ Consiglio Nazionale delle Ricerche}}\emph{,}
\textit{\emph{IEIIT}} institute, Piazza Leonardo da Vinci 32, I-20133,
Milano, Italy.\linebreak{}
 }}
\maketitle
\begin{abstract}

Recently, proposals of human-sensing-based services for cellular and local area networks have brought indoor localization to the attention of several research groups. In response to these stimuli, various Device-Free Localization (DFL) techniques, also known as Passive Localization methods, have emerged by exploiting ambient signals to locate and track individuals that do not carry any electronic device. This study delves into human passive indoor localization through full-wave electromagnetic simulations. For the scope, we exploit simulations from the commercial tool FEKO\textsuperscript{\textregistered} software that employs the Method of Moments (MoM). In particular, we collect and analyze the electric field values in a scenario constituted by a dense 2D/3D deployment of receivers in the presence of an anthropomorphic mobile target. The paper describes in detail the collected dataset and provides a first analysis based on a statistical approach. Possible use cases are also investigated through examples in the context of passive localization, sensing, and imaging. 

\end{abstract}


\section{Introduction}
\label{sec:intro}

\IEEEPARstart{P}{assive} radio sensing utilizes stray ambient radio signals emitted by Radio Frequency (RF) devices to identify, locate, and monitor individuals not carrying any electronic devices, commonly referred to as device-free \cite{youssef-2007,savazzi-2016,nuzzer}. By leveraging the \textit{Integrated Sensing and Communication} paradigm, DFL networks utilize RF nodes of cellular or wireless area networks to measure and evaluate the increased attenuation caused by human bodies within the monitored area, enabling the estimation of their positions and movements. In fact, it is well known that the presence or the movements of people induce alterations of the EM field \cite{koutatis-2010,krupka-1968} collected by a wireless device. Thus, radio signals inherently capture electromagnetic (EM) information about all static or moving objects encountered during their propagation.

To retrieve these information, different kinds of localization algorithms have been investigated for DFL systems \cite{savazzi-2016}, such as compressive sensing \cite{wang-2012}, Machine/Deep Learning (ML/DL) techniques \cite{shit-2019,palip,santoboni-2022}, RTI (Radio Tomographic Imaging) \cite{wilson-2010}, and Bayesian algorithms \cite{rampa-2021,kat}. 
Proposed solutions to the radio sensing problem necessitate an empirical \cite{{wilson-2010}}, statistical \cite{kat}, or physical \cite{rampa-2017,rampa-2022a} model to interpret how human activities influence radio wave propagation. The influence of body movements on radio signals can be predicted through considerations rooted in electromagnetic (EM) propagation theory \cite{krupka-1968}. Recently, numerous physical and statistical models have been introduced, exploiting various methodologies such as ray tracing \cite{phyindoor,rayback}, moving point scattering \cite{scatt}, Geometrical Theory of Diffraction (GTD) \cite{james-2007,qi-2017}, Uniform Theory of Diffraction (UTD) \cite{plouhinec-2023,qi-2017}, and scalar diffraction theory \cite{koutatis-2010,rampa-2022a,j-fieramosca-2023}. However, existing full-wave models are often deemed overly intricate and lengthy for practical application in real-time sensing scenarios \cite{eleryan-2011}. To strongly reduce the simulation time, Generative Neural Networks (GNNs) have been recently emerged as promising tools for fast field computation and to solve inverse imaging problems \cite{generative_mag} as well when properly trained with off-line simulation results \cite{generation}. 

\begin{figure}
\begin{centering}
\includegraphics[scale=0.365]{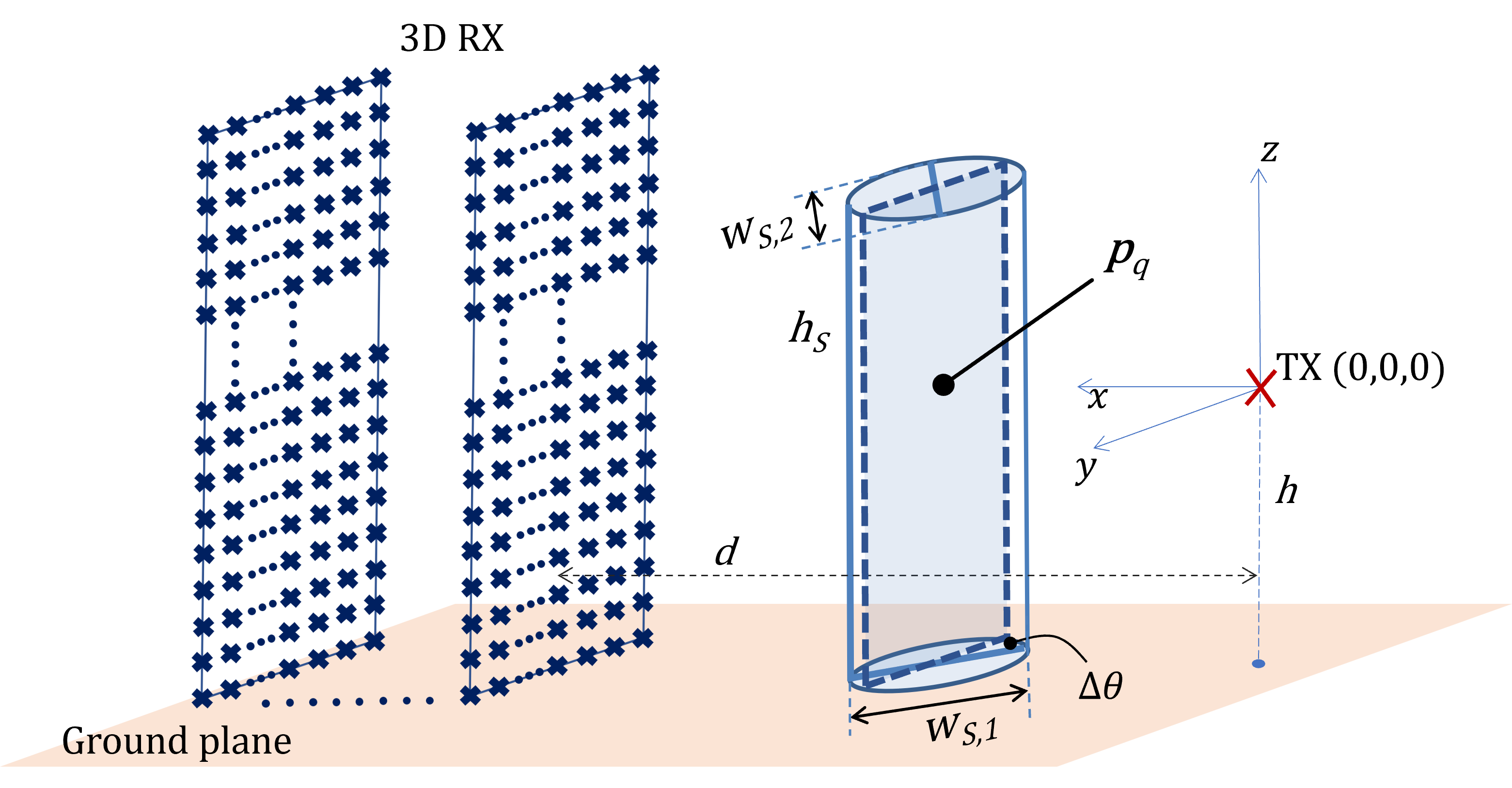}
\par\end{centering}
\caption{\label{fig:scenario}3-D deployment of the radio link using a 3D array of $50\times90\times180$ measurement points. The center of the first surface of the array is placed at distance $d=4$ m from the source (TX) and it is orthogonal to the Line-Of-Sight (LOS) path connecting the TX  with the measurement points (3D RX). The body is sketched as an homogeneous perfectly absorbing 3D elliptical cylinder of height $h_S$ with maximum $W_{S,1}$ and minimum $W_{S,2}$ traversal size.}
\end{figure}

\textbf{Contributions.} The paper describes a novel dataset obtained with the full-wave EM software simulator FEKO\textsuperscript{\textregistered} \cite{jakobus-2012}. The simulations are intended to support the analysis of the human body blockage effects on the EM field observed by receivers organized in dense 2D and 3D array layouts. A realistic body-shaped target is also simulated using the approach shown in \cite{bartlet}. Preliminary data analysis is also shown. We discuss: 
\begin{itemize}
    \item a simple approach to body imaging obtained by purely statistical considerations and based on mean and standard deviation analysis of the received EM field samples;
    \item an analysis of the effects of small, or involuntary, movements of the simulated body on the EM field samples for selected receivers from 1D/2D/3D layouts.
\end{itemize}
The dataset and simulation results that are discussed in this paper are available online in \cite{dataport} and can be used to carry out research in the field of EM body modelling, array and beamforming processing \cite{mimo}, and physics-driven machine learning \cite{generation} as well. 

The paper is organized as follows. Sect.~\ref{sec:setup} introduces the problem and the setup adopted for simulations, while Sect.~\ref{sec:mom} discusses the basics of the Method of Moments (MoM), used by the EM simulator FEKO\textsuperscript{\textregistered}. Sect.~\ref{sec:simulations} describes how the simulations are performed with FEKO\textsuperscript{\textregistered}, while Sect.~\ref{sec:analysis} presents a first analysis of the data both from an imaging and a statistical perspective. Concluding remarks and future research are finally discussed in Sect.~\ref{sec:conclusions}.

\begin{center}
\begin{figure}
\begin{centering}
\includegraphics[scale=0.6,center]{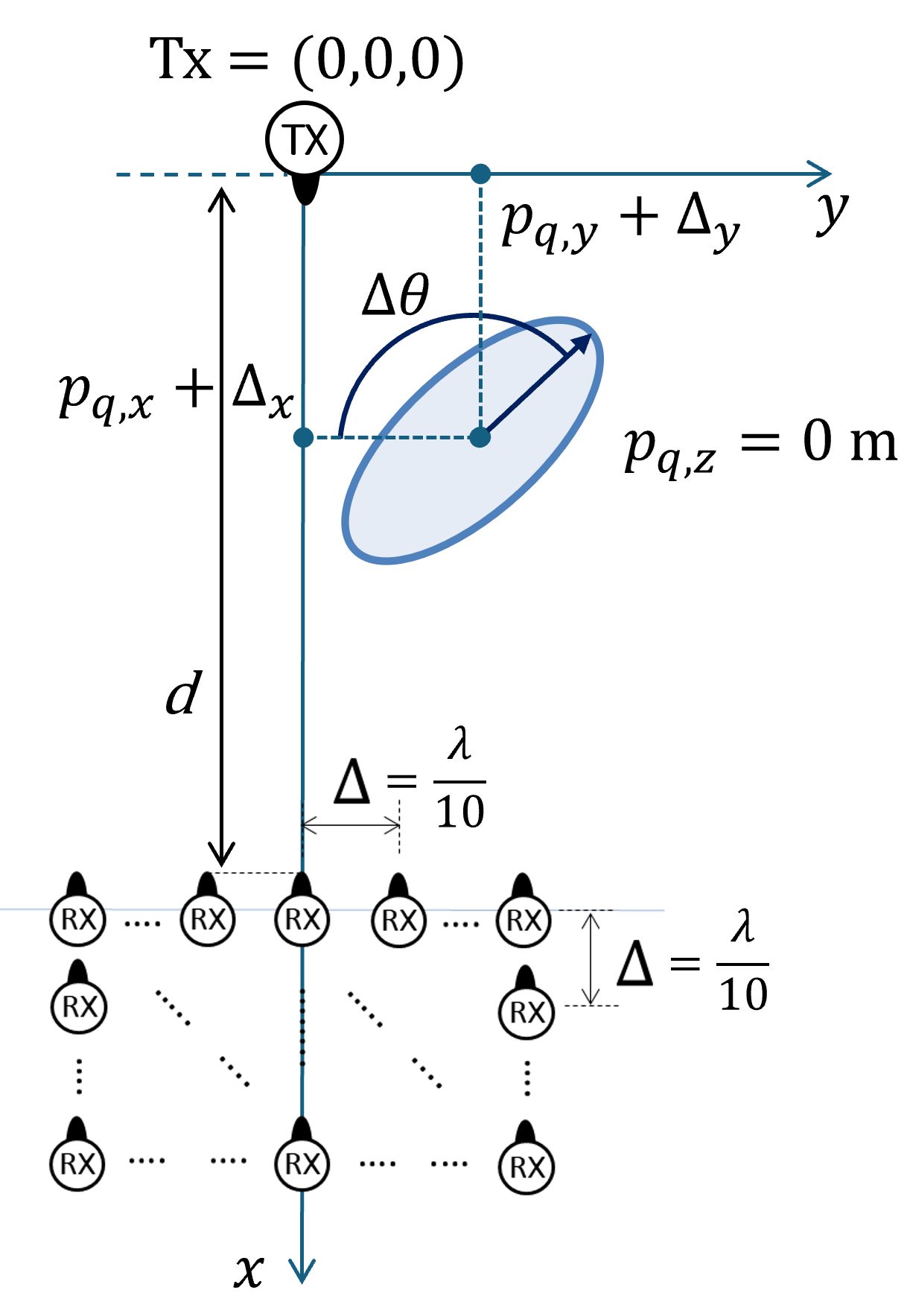}
\end{centering}
\caption{\label{fig:2d-layout} Horizontal 2-D layout of the radio links corresponding to Fig.~\ref{fig:scenario}.}
\end{figure}
\end{center}

\begin{center}
\begin{figure}
\begin{centering}
\includegraphics[scale=0.64,center]{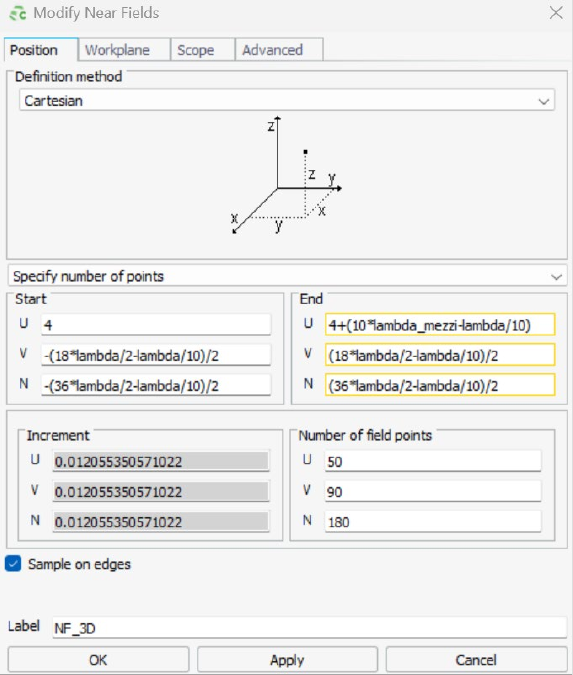}
\end{centering}
\caption{\label{fig:feko} FEKO\textsuperscript{\textregistered} simulation settings for the measurement array.}
\end{figure}
\end{center}

\section{Simulation scenario}
\label{sec:setup}

The simulation scenario involves an Hertzian dipole radiating at the nominal frequency $f = 2.4868$ GHz and located at $h=0.99$ m from the floor. The radiated field is measured over an array of points organized in a $3$D layout as depicted in Fig.~\ref{fig:scenario}. This comprises multiple 2D arrays that are uniformly deployed on co-located surfaces positioned at varying distances from the source. Specifically, the first measurement surface, i.e. the most close to the source, is located $4$ m away (distance $d=4$ m) from the source itself (and its center is aligned with the source), and it is followed by other $49$ additional surfaces spaced at intervals of $\Delta=\frac{\lambda}{10}$. Moreover, each surface is composed of a grid of $90\times180$ elements, also spaced $\Delta=\frac{\lambda}{10}$ apart. Fig.~\ref{fig:feko} shows how these parameters have been set on FEKO\textsuperscript{\textregistered}. Each element of the 3D-array is then equally spaced of $\Delta=\frac{\lambda}{10}$ with respect to the adjacent elements both in vertical and horizontal directions, as shown in Figs.~\ref{fig:scenario} and \ref{fig:2d-layout}, respectively.

\begin{center}
\begin{figure}
\begin{centering}
\includegraphics[scale=0.54,center]{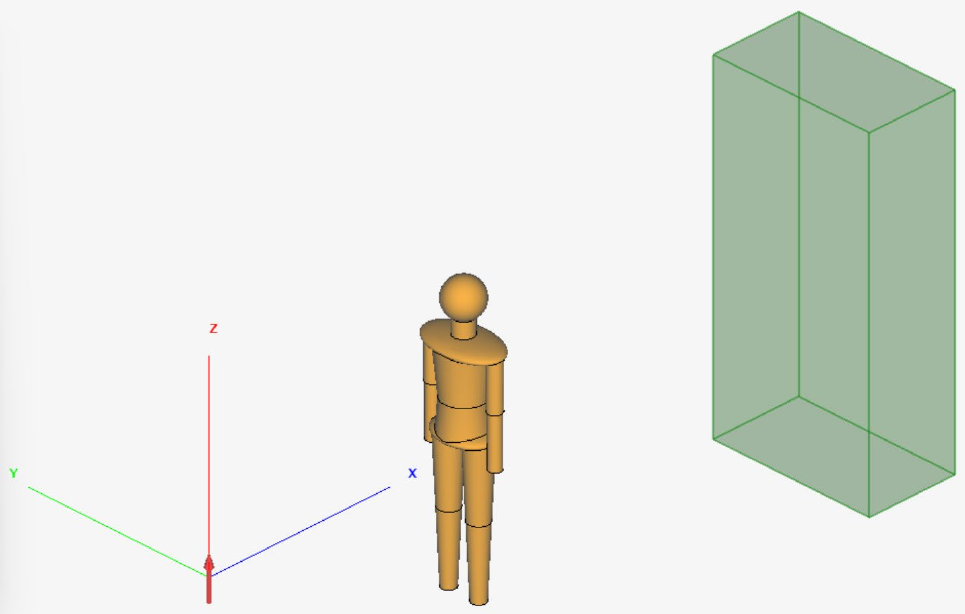}
\end{centering}
\caption{\label{fig:anto} Body-shape EM model designed with FEKO\textsuperscript{\textregistered}. The green rectangular base prism on the right represents the bounding box of the set of the measurement points.}
\end{figure}
\end{center}

\vskip -9mm

In Figs.~\ref{fig:scenario} and \ref{fig:2d-layout}, the body is sketched as an homogeneous perfectly absorbing 3D elliptical cylinder. This is the model that, by exploiting the scalar diffraction theory, has been employed in the previous papers \cite{rampa-2021,rampa-2017,j-fieramosca-2023,fieramosca-2023} to evaluate the attenuation introduced by the body in the radio link area. However, for more realistic FEKO\textsuperscript{\textregistered} simulations, we exploit here a more detailed anthropomorphic target. The proposed body-shaped EM model \cite{fieramosca-2024} is designed as shown in Fig.~\ref{fig:anto}, featuring an height $h_S$ of $1.80$ m, a maximum transversal width $w_{S,1}$ of $0.52$ m, and a maximum orthogonal thickness $w_{S,2}$ of $0.32$ m. According to the considered layout, each individual surface fully encloses the shape of the considered body. The body material properties are obtained from a simulator available online \cite{bartlet}. We assume that the body is primarily composed of muscles having relative permittivity $\epsilon_r$ set to $60$, dielectric loss tangent tan$\delta$ set to $0.242$, and mass density $\rho$ set to $1040.0$ kg/$m^3$. This FEKO\textsuperscript{\textregistered} model represents an improvement over the previous built-in $2$D Perfect Electric Conductor (PEC) model used to compare the results obtained with the simplified model adopted in \cite{rampa-2021,rampa-2017,j-fieramosca-2023} against the FEKO\textsuperscript{\textregistered} simulations. 

\section{Method of moments}
\label{sec:mom}

The default solver within FEKO\textsuperscript{\textregistered} is the Method of Moments (MoM), one of the most important general methods for solving electromagnetic-field problems. MoM operates on the principle of discretizing the geometry of the structure into small elements, typically triangles for surface discretization \cite{davidson}. Appropriate basis functions are then chosen, considering continuity and boundary conditions, in order to represent the unknown current distribution on each element. Assembling the contributions from all elements, a linear system of equations in matrix form is set and then solved to determine the unknown currents. Generally, matrix dimensions range from hundreds (small antenna problems) to several thousand, although  computationally expensive \cite{harrington}.  

One of the well-known weakness of MoM is that it does not handle very well penetrable materials, since they require fictitious equivalent volumetric currents that are very expensive computationally \cite{davidson}.
However, the MoM implementation in FEKO\textsuperscript{\textregistered} is enriched by the use of the Surface Equivalent Principle (SEP) \cite{li-2007}. The SEP is a foundational concept that simplifies the representation of dielectric bodies within the simulation environment: when a closed dielectric body is encountered in the simulation, SEP introduces equivalent electric and magnetic currents along the surface of the body (opposed to using only equivalent electric currents on a perfectly electric conducting body). The SEP can handle homogeneous dielectric bodies (as in our case) efficiently; however FEKO\textsuperscript{\textregistered} includes other solution methods for efficient treatment of inhomogeneous dielectric structures that can be further investigated in future works.

\section{Simulations of body movements} 
\label{sec:simulations}

FEKO\textsuperscript{\textregistered} simulations were conducted with the body positioned at various distances and orientations relative to the source and the measurement points, allowing the collection of both real and imaginary samples of the  EM field at the receivers' location. For the sake of simplicity, no other obstacles, walls, floor, and ceiling are included in the current simulations. However, they can be easily added in the simulated scenario at the cost of an increased processing time.

In particular, considering the 2D array placed in the vertical plane $x=d=4$ m, the simulations involve 3 distinct body positions $\mathbf{p}_{q}=(p_{q,x},p_{q,y},p_{q,z}$) with $q=1,2,3$ while the corresponding coordinates are given by: 
\begin{enumerate}
    \item $\mathbf{p}_1$: $p_{1,x}=2$ m, $p_{1,y}=0$ m, $p_{1,z}=0$ m,
    \item $\mathbf{p}_2$: $p_{2,x}=2$ m, $p_{2,y}=0.25$ m, $p_{2,z}=0$ m,
    \item $\mathbf{p}_3$: $p_{3,x}=2$ m, $p_{3,y}=0.50$ m, $p_{3,z}=0$ m.
\end{enumerate}
Notice that the third coordinate is always set to zero. In fact, the center of the body is always located at $z=0$, making its distance from the ground plane equal to $h=0.99$ m, the same of the transmitting antenna. Since the $z$ coordinate does not change, in the following discussion we will omit it.

Further simulations were conducted to emulate the inherent micro-movements observed in real-life scenarios \cite{rampa-2017}, such as bodily motions. The simulations consider only small translations along both $x$ and $y$ axes (i.e., no vertical movements along the $z$ axis are allowed), with increments of $\pm \frac{\lambda}{4}$ w.r.t. the positions $\mathbf{p}_{q}$. Therefore, these micro-movements imply different body positions $\mathbf{p}'_q = (p_{q,x} +\Delta_x, p_{q,y} + \Delta_y, p_{q,z})$ with $\Delta_x = \Delta_y = \pm \frac{\lambda}{4}$ and $p_{q,z}=0$. 
Different orientations of the body (rotational movements) were also simulated to assess their impact on imaging/positioning accuracy. Body rotations were performed always at each target position, considering angles/orientations of $\Delta\theta\in\{\Delta\theta_1,\Delta\theta_2,\Delta\theta_3,\Delta\theta_4\}$ w.r.t. the default target orientation with $\Delta\theta_1 =0^{\circ}$, $\Delta\theta_2 =45^{\circ}$, $\Delta\theta_3 =90^{\circ}$ and $\Delta\theta_4 =135^{\circ}$. These micro-movements transformations aimed to mimic involuntary motions or instabilities inherent to such motions. By incorporating these variations into the simulation framework, the proposed dataset is designed to be representative of a realistic body motion scenario.

\section{Data analysis} 
\label{sec:analysis}
In what follows, as an application example, the simulated data are exploited to generate 2D maps of the attenuation induced by the body on each vertical 2D surface.

\subsection{Imaging results}
We obtain a measure of the body-induced attenuation associated to
each spatial element $i,j$ of the 2D surface (lying on the $xy$ plane with $z=0$) with respect to different
(nominal) macro-positions $\mathbf{p}_{q}$ and by averaging over
different body positions $\mathbf{p}'_{q}$ due to the micro-movements, namely the roto-translations due to $\Delta_{x},\Delta_{y}$, and $\Delta\theta$
w.r.t. the nominal positions and orientations. For each 2D receiver array element,
the corresponding attenuation elements can be computed as:

\begin{equation}
\Bar{H}_{i,j}=\int_{S_{x,y}}h_{i,j}(p_{q,x}+x,p_{q,y}+y)\,f(x,y)\,dx\,dy
\label{eq:average}
\end{equation}
where $h_{i,j}(x,y)$ is the RF attenuation (dB) obtained as: 

\begin{equation}
    h_{i,j}(x,y)=-10\,\log_{10}\left|E_{i,j}(x,y)\,/\,E_{i,j}(0)\right|^{2}
\end{equation}
being $E_{i,j}(x,y)$ the received electric field associated to the spatial element $i,j$ of the 2D array in the presence of the body at nominal position $\mathbf{p}_q$ and with respect each micro-movement $x,y$ and $E_{i,j}(0)$ being the corresponding field in absence of the target. $f(x,y)$ measure the probability of observing each micro-movement $x,y$ inside the surface $S_{x,y}$ having area $\lambda^{2}/4$. A uniform probability measure is assumed in the following. The integral in (\ref{eq:average}) is also replaced with the sum over $36$ discrete micro-movements as described in Sect.~\ref{sec:simulations}. 

As shown for the average attenuation, a map of the standard deviations
$\Delta{H}_{i,j}$ is constructed for each 2D array element. In particular,
the standard deviation sample associated to the measurement point $i,j$
is calculated with respect to the corresponding mean attenuation value
and relative to the same micro-movements and rotations patterns as: 
\begin{equation}
\Delta{H}_{i,j}^{2}=\int_{S_{x,y}}[h_{i,j}(p_{q,x}+x,p_{q,y}+y)-\Bar{H}_{i,j}]^{2}\,f(x,y)\,dx\,dy.
\end{equation}

The resulting image maps of average $\{\Bar{H}_{i,j}\}_{i=1,j=1}^{90, 180}$ and deviation patterns $\{\Delta{H}_{i,j}\}_{i=1,j=1}^{90, 180}$ can be further processed to provide insights into the impact of the anthropomorphic obstacle and its movements on propagation in the simulated environment. 

\begin{center}
\begin{figure}
\begin{centering}
\includegraphics[width=0.9\columnwidth,center]{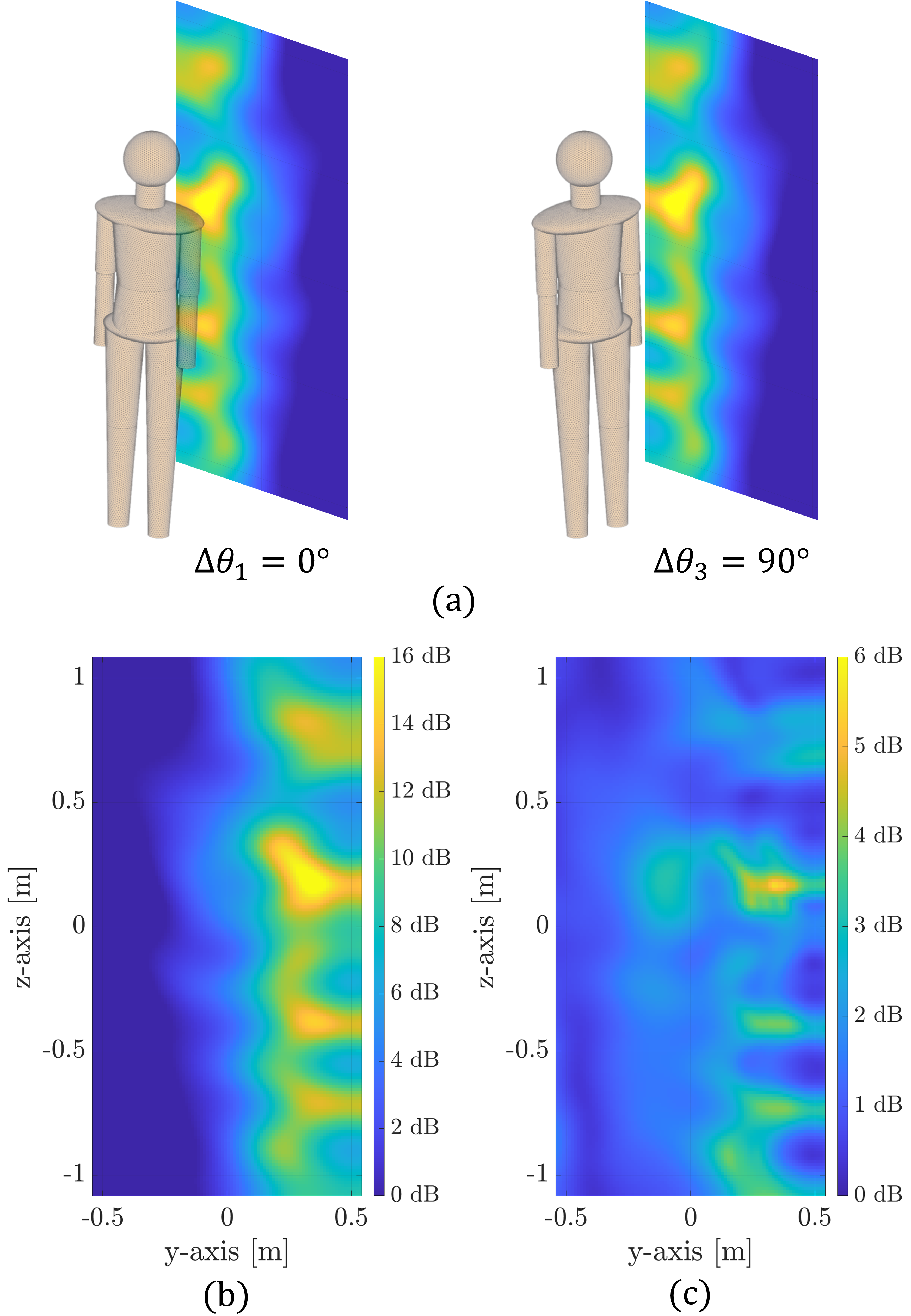}
\par\end{centering}
\caption{\label{fig:mappe} Imaging maps of attenuation $\Bar{H}_{i,j}$ observed on a 2D surface grid of $90\times180$ elements. (a) Positions and orientations of the simulated body highlighted; (b) average attenuation map $\Bar{H}_{i,j}$; and (c) corresponding standard deviation map $\Delta{H}_{i,j}$.}
\end{figure}
\par\end{center}

\begin{center}
\begin{figure}
\begin{centering}
\includegraphics[width=0.9\columnwidth]{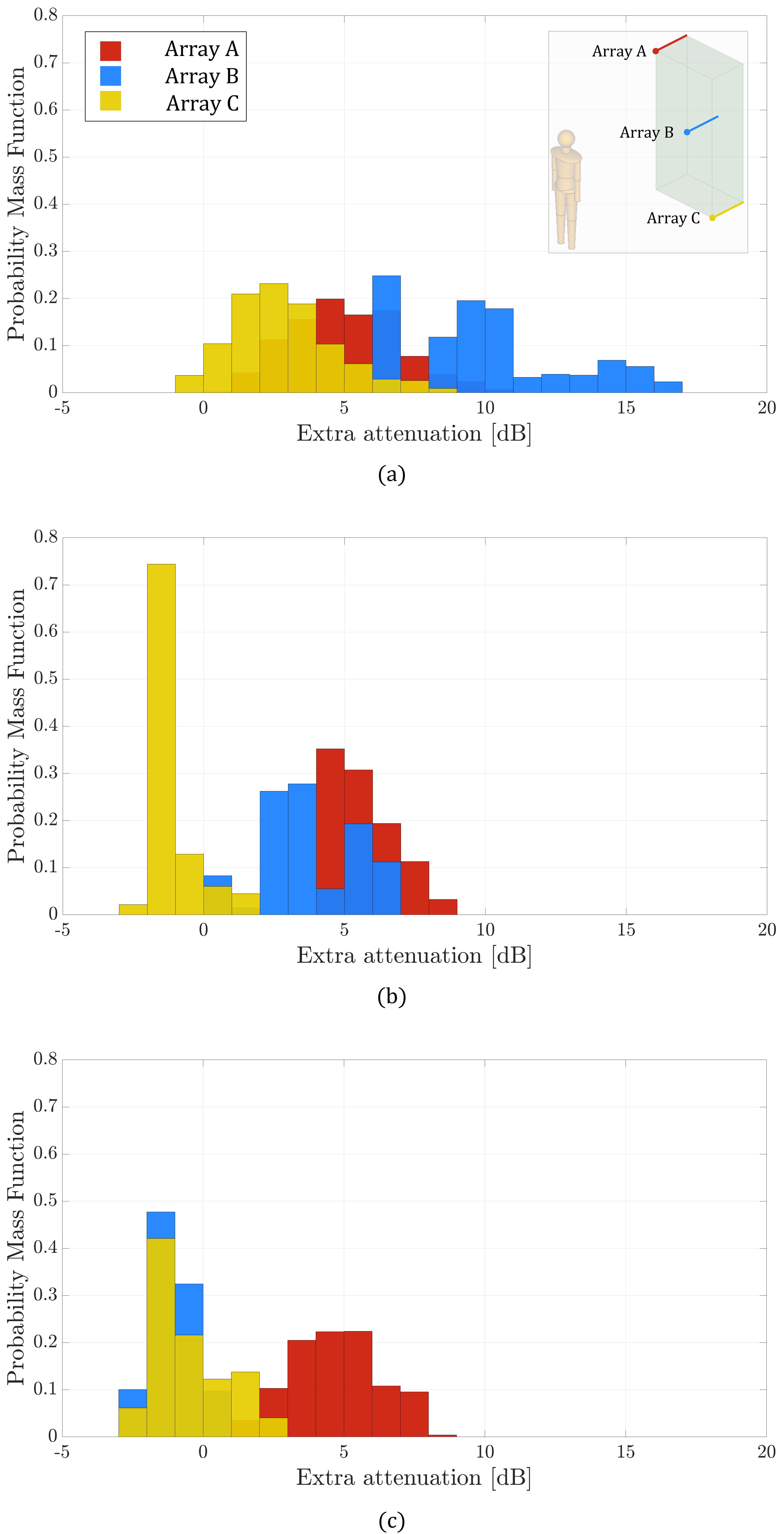}
\end{centering}
\caption{\label{fig:pdfs}Probability distribution of the observed EM field attenuation w.r.t. body movements in 3 selected nominal locations $\textbf{p}_1$, $\textbf{p}_2$, and $\textbf{p}_3$. Data is collected from receivers which belong to three 1D arrays parallel to the LOS and located in A, B and C as sketched in the top right part of (a). Array A is located at position $x=0$ m, $y=0.536$ m, and $z=1.07$ m on the grid, array B at $x=0$ m, $y=0$ m, and $z=0$ m, while array C at $x=0$ m, $y=-0.536$ m, and $z=-1.07$ m. From top to bottom: (a) target in position $\textbf{p}_1$; (b) target in position $\textbf{p}_2$; and (c) target in $\textbf{p}_3$.}
\end{figure}
\end{center}

Fig.~\ref{fig:mappe} reports an example of average and deviation maps for the considered scenario and 2D measurement array. The body is located in position $\textbf{p}_2$ ($p_{2,x}=0$ m, $p_{2,y}=0.25$ m). Fig.~\ref{fig:mappe}(a) further highlights the overlapping between the average attenuation map and the body rotated by each $\Delta\theta$ to help relate the final pattern with some of the movements considered. Analyzing both maps, 
the body footprint is clearly distinguishable as well as parts of the torso and head. The regions with highest attenuation are those consistent over translations and rotations of the body. When the target moves, only a subset of receivers are affected by such movements. Using the relative position of these receivers in the 2D array manifold, it is possible to trace back the offset of the body blockage from the LOS path. 
Also, by observing the corresponding standard deviation maps, it is possible, for example, to distinguish specific human body parts most susceptible to movements and variability, i.e., the arm movements and shape.  

\subsection{Statistical processing and impact of small body movements}
Figure $\ref{fig:pdfs}$ shows the probability distribution (probability mass function) of the observed EM field attenuation relative to three groups of 50 points each, which belong to three 1D (i.e., linear) arrays parallel to the LOS and located in grid positions A, B and C identified by the same coordinate $x=0$ while $y$ and $z$ coordinates are set as:
\begin{itemize}
    \item A: $y=+45\cdot\frac{\lambda}{10}=0.536$ m, $z=+90\cdot\frac{\lambda}{10}=1.07$ m;
    \item B: $y=0$ m, $z=0$ m;
    \item C: $y=-45\cdot\frac{\lambda}{10}=-0.536$ m, $z=-90\cdot\frac{\lambda}{10}=-1.07$ m.
\end{itemize} 
Probability functions are obtained for three different nominal positions of the body blockage in $\textbf{p}_1$, $\textbf{p}_2$ and $\textbf{p}_3$ (see Sect.~\ref{sec:simulations}), as depicted in Fig.~\ref{fig:pdfs} (a), (b), and (c), respectively. For each case, we used  $36\times50$ attenuation samples, namely $50$ measurements of $36$ micro-movements.

The goal of the analysis is to evaluate and discern the relative positions of the receiver arrays on the 2D grid that are most sensitive to body blockage movements. 
For instance, we observe that the distribution associated with the receiver array A does not exhibit consistent changes in behavior from $\textbf{p}_1$ to $\textbf{p}_3$. These receivers are situated on an area of the 2D grid where the body blockage remains consistently visible, resulting in detected attenuation consistently ranging between $0-10$ dB. In contrast, receiver array C ceases to detect the body at $\textbf{p}_2$, with the attenuation curve never exceeding the $2$dB. Receivers that belong to array B are located in the center of the $2$D grid and demonstrate the capability to follow target movements. When the body is along the LOS, the attenuation curve averages around $10$ dB and can even peak at $15$ dB. However, as the target moves, the curve's mean decreases as the subject is no more visible.

\section{Conclusions and future activities}
\label{sec:conclusions}

The paper describes a simulation dataset which supports the analysis of the human body blockage effects on the EM field observed by receiver antennas organized in large and dense 2D and 3D array layouts. A simple tool for body imaging was developed using statistical considerations. The tool is based on the analysis of mean and standard deviation of the simulated RF attenuation samples.
Furthermore, an investigation was conducted into the impact of involuntary movements of the simulated body. 

The dataset is intended to support the analysis and the validation of array processing techniques for passive indoor localization such as Direction of Arrival estimation (DOA) on azimuth/elevation, which tracks the perturbations of the DOA field as induced by body movements as well as 2D beamforming. 
This constitutes an advancement from prior studies \cite{mimo} conducted using the linear beamforming approach.

Future works will be focused on the design and implementation of array processing algorithms to detect variations in target height/shape and detect movements of body parts, with application to fall detection \cite{kianoush-2016} and activity recognition \cite{palip}.
Furthermore, the study can be extended to tomographic analysis, facilitating a deeper insight into the spatial distribution of signals over 2D and 3D dimensions. 

Furthermore, considering passive body localization applications, one aspect that requires particular attention is the study of the impact of body micro-movements on the magnitude/phase of EM field and the points of the 1D/2D/3D array manifold that are most vulnerable to them. The analysis is of interest for holography \cite{holl-2017}, ray-tracing \cite{na-2024}, tomography \cite{hamilton-2014,tomog}, and array processing \cite{santoboni-2022} scenarios. 


\end{document}